\documentclass[twocolumn, superscriptaddress]{revtex4-2} 
\usepackage[utf8]{inputenc}
\usepackage[a4paper, left=15mm, right=15mm, top=25mm, bottom=25mm]{geometry}
\usepackage{graphicx}
\usepackage{siunitx}
\usepackage{amssymb}
\usepackage{amsmath}
\usepackage{appendix}
\usepackage[dvipsnames]{xcolor}
\usepackage[normalem]{ulem}
\usepackage{braket}
\usepackage{orcidlink}

\begin{document}
\title{An ionic clock qubit inside a circular Rydberg atom}

\author{Fabian Thielemann \orcidlink{https://orcid.org/0000-0003-4578-8500}}
\affiliation{5. Physikalisches Institut and Center for Integrated Quantum Science and Technology,
	Universität Stuttgart, Pfaffenwaldring 57, 70569 Stuttgart, Germany}
\author{Aaron Götzelmann\orcidlink{https://orcid.org/0000-0001-5527-5878}}
\affiliation{5. Physikalisches Institut and Center for Integrated Quantum Science and Technology,
	Universität Stuttgart, Pfaffenwaldring 57, 70569 Stuttgart, Germany}
\author{Marius Thomas \orcidlink{https://orcid.org/0009-0002-4096-6101}}
\affiliation{5. Physikalisches Institut and Center for Integrated Quantum Science and Technology,
	Universität Stuttgart, Pfaffenwaldring 57, 70569 Stuttgart, Germany}
\author{Einius Pultinevicius \orcidlink{https://orcid.org/0009-0005-7404-9178}}
\affiliation{5. Physikalisches Institut and Center for Integrated Quantum Science and Technology,
	Universität Stuttgart, Pfaffenwaldring 57, 70569 Stuttgart, Germany}
\author{Armin Humić \orcidlink{https://orcid.org/0009-0009-4628-7178}}
\affiliation{5. Physikalisches Institut and Center for Integrated Quantum Science and Technology,
	Universität Stuttgart, Pfaffenwaldring 57, 70569 Stuttgart, Germany}
\author{Christian Hölzl \orcidlink{https://orcid.org/0000-0002-2176-1031}}
\affiliation{5. Physikalisches Institut and Center for Integrated Quantum Science and Technology,
	Universität Stuttgart, Pfaffenwaldring 57, 70569 Stuttgart, Germany}
\author{Florian Meinert \orcidlink{https://orcid.org/0000-0002-9106-3001}}
\affiliation{5. Physikalisches Institut and Center for Integrated Quantum Science and Technology,
	Universität Stuttgart, Pfaffenwaldring 57, 70569 Stuttgart, Germany}

\date{\today}

\begin{abstract}
	Neutral atoms trapped in optical tweezers and excited to Rydberg states, together with trapped ions, are among the most advanced platforms
	for quantum simulation and quantum computing. Current experiments often rely on additional atoms in neighboring traps to encode ancilla qubits for local
	manipulation and readout. Here, we demonstrate a dual ion–Rydberg system comprising two qubits encoded in two individually controlled
	electrons of the same alkaline-earth atom. The first, a microwave qubit, is encoded in a pair of circular Rydberg states, while the second, an optical qubit,
	is encoded on a narrow quadrupole transition of the Rydberg atom's ionic core. We demonstrate coherent control of the optical qubit and achieve coherence
	times of several hundred microseconds under dynamical decoupling. Furthermore, we realize coherent coupling between the two electrons via
	electrostatic quadrupole interactions over the large separation between the Rydberg electron and the ionic core, and map out
	its angular tunability. Finally, we demonstrate a two-qubit operation, reminiscent of a Mølmer–Sørensen gate, that evolves through
	an entangled state of the two qubits driven by the quadrupole coupling. Our work opens a pathway to exploit a pair of individually controlled
	electronic qubits with tunable coupling for quantum simulation and quantum metrology.
\end{abstract}

\maketitle

\section{Introduction}

Quantum correlations and entanglement are ubiquitous among bound electrons in
atoms. Both emerge from exchange interactions resulting from the overlap of
electronic orbitals and underpin the atomic level structures observed
in nature \cite{Heisenberg1926,TannerHeliumReview2000}. Exploiting multiple
electrons within a single atom as independently controllable coherent resources
for quantum information processing, however, is typically hindered by these
same interactions. Although spectroscopy schemes involving more than a single
electron are powerful tools in precision measurements and metrology
\cite{SafronovaTwoClockTransitions2018,DzubaStandardModelTests2018,IshiyamaInnerShellClockTransition2023},
electronic correlations naturally prevent the manipulation of one electron
without affecting the other. Moreover, atomic states with two excited valence
electrons are typically unstable and short-lived when autoionization, a decay
process driven by the mutual interactions, is energetically allowed
\cite{fanoEffectsConfigurationInteraction1961,ArimondoAutoionizationReview2010}.

Even if one of the electrons is prepared in a highly excited Rydberg state
with large principal quantum number $n$, finite wavefunction overlap with the
second low-$n$ electron still leads to rapid ionization
\cite{LocheadIonizationBulkGas2013,mcquillenImagingEvolutionUltracold2013,LehecIsolatedCoreExcitation2021}.
This situation changes dramatically for circular Rydberg states with maximally
allowed angular momentum ($m=l=n-1$, where $m$ and $l$ denote the magnetic and
orbital angular momentum quantum numbers, respectively)
\cite{KleppnerCircularStates1983,RaimondCRSReview2001}. In these states, the
electronic wavefunction is localized in a narrow torus with negligible overlap
with low-lying core orbitals, strongly suppressing autoionization and
exchange coupling
\cite{rousselObservationCircularMetastableDoubly1990,TeixeiraNonAutoionizingRydberg2020}.

For a two-electron atom such as strontium, this enables coherent manipulation
of the optically active core electron, allowing the internal degrees of freedom
of the embedded Sr$^+$ ionic core to be harnessed as an independently
controllable quantum degree of freedom
\cite{leibfriedQuantumDynamicsSingle2003}. Moreover, long-ranged electrostatic
interactions between the electrons remains significant, even if the circular
electron orbits at distances nearly three orders of magnitude larger than the
characteristic size of the low-lying core orbitals
\cite{muniOpticalCoherentManipulation2022,wirthQuadrupoleCouplingCircular2024}.
This allows for implementing switchable and tunable coupling between the ionic
core and the Rydberg electron, enabling the controlled generation of electron
correlations and entanglement.

\begin{figure*}[t]
	\centering
	\includegraphics[width=\textwidth]{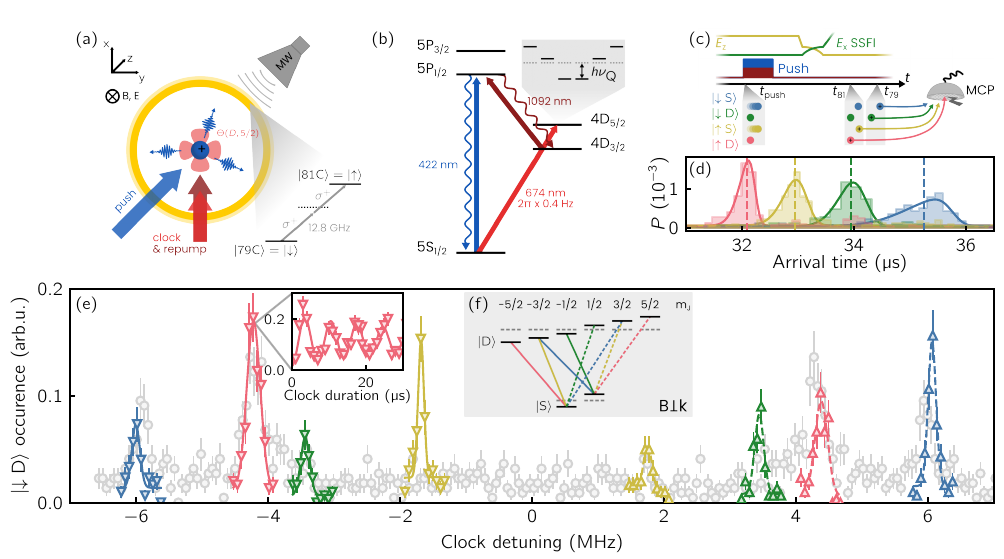}
	\caption{\textbf{Manipulation and detection of the clock qubit inside a circular Rydberg atom.}
		(a) We excite one of the two valence electrons of $^{88}$Sr to the circular
		Rydberg state $\ket{79\text{C}}$ (yellow). A two-photon microwave transition to the
		$\ket{81\text{C}}$ state allows us to coherently control the state of the
		Rydberg qubit. We use a pair of push and repump beams to rapidly scatter
		photons on the ionic core D1 transition. Our clock laser, propagating orthogonal to the magnetic
		field $\mathbf{B}$, excites the ionic core to the D$_{5/2}$ state (colors used for the different beams match those used in b).
		(b) Level scheme of the $^{88}$Sr$^+$ ionic core: The dipole allowed S$_{1/2}\rightarrow$P$_{1/2}$
		(D$_{3/2}\rightarrow$P$_{1/2}$) transition is used for pushing (repumping), while the
		S$_{1/2}\rightarrow$D$_{5/2}$ quadrupole transition encodes an optical qubit. In the presence of
		a quadrupole field, the magnetic sublevels of the D$_{5/2}$ state experience an energy shift
		$h\nu_\text{Q} \propto m_J^2$. (c) Two-qubit
		readout scheme: a push sequence at $t_\text{push}$ displaces atoms in the S$_{1/2}$ state, while those shelved to the
		D$_{5/2}$ state remain in place. The electric field is subsequently rotated from $z$ to $y$ and
		ramped up for SSFI, ionizing the atoms at $t_{81}$ and $t_{79}$,
		depending on their Rydberg state. As the atoms displaced towards the microchannel-plate detector are accelerated over a shorter distance,
		they arrive later. This leads to four distinct peaks in the arrival
		time histogram, shown in (d), corresponding to the populations of the four basis states in the two-qubit system: $\ket{\downarrow\text{S}}$ (blue), $\ket{\downarrow\text{D}}$ (green), $\ket{\uparrow\text{S}}$ (yellow) and $\ket{\uparrow\text{D}}$ (red).
		(e) Spectroscopy of the clock transition: to record this spectrum the Rydberg electron was prepared in the
		$\ket{79\text{C}}$ state. We then applied a shelving pulse on the clock transition, and increased the push duration
		such that all atoms in the S$_{1/2}$ state were lost. We set
		$\mathbf{B}\perp \mathbf{k}$, where $\mathbf{k}$ is the
		$k$-vector of the optical clock laser, and rotate the polarization such
		that we can address all transitions with $ \Delta m = \pm1, \pm2$ as shown
		in (f). Transitions with negative (positive) relative magnetic moment
		$\mu_\text{rel}$ are indicated with solid (dashed) lines. A wide scan with
		$\text{pol}\perp \mathbf{B}$ ($\Delta m=\pm2$) is shown in grey. All errorbars
		represent $1\sigma$ confidence intervals.}
	\label{fig:fig1}
\end{figure*}

Here, we demonstrate such a two-electron atomic system in which each electron
encodes a long-lived qubit that can be coherently controlled and manipulated
(Fig.~\ref{fig:fig1}a
and b). Using individual strontium atoms trapped in optical tweezers
\cite{cooperAlkalineEarthAtomsOptical2018,norciaMicroscopicControlDetection2018a},
we encode the first, microwave-controlled qubit in a pair of circular Rydberg
orbits ($n=79$ and $n=81$) with millisecond-scale lifetimes
\cite{holzlLongLivedCircularRydberg2024a,pultineviciusLonglivedGiantCircular2025a}.
The second qubit is realized on the Sr$^+$ 5S$_{1/2} \rightarrow 4$D$_{5/2}$ ultra-narrow
optical quadrupole transition \cite{roos2000controllingIons}, which serves as
the clock transition in trapped-ion atomic clocks and as an optical qubit in
trapped-ion quantum computing
\cite{akermanOperatingMultiIonClock2025,ManovitzTrappedIonQC2022,lindvall88SrOpticalClock2025}.
In our experiments, we demonstrate coherent control of the optical ion qubit
within the large circular Rydberg orbit, as well as simultaneous control on both
qubits over timescales of hundreds of microseconds.

We characterize the electrostatic quadrupole interaction between an ionic qubit
and a Rydberg qubit, and demonstrate its tunability via the relative orientation
of the ionic D-orbital and the circular Rydberg orbit. This
gives rise to a “magic angle” at which the coupling vanishes. The
electron–electron interaction further imprints a two-qubit-state-dependent
phase shift, forming the basis for quantum logic spectroscopy of a circular
microwave qubit read out via a laser-addressable ionic clock transition.
Finally, we implement a gate sequence analogous to a M{\o}lmer–S{\o}rensen-type
interaction, realizing two-qubit state rotation and transiently generating a
Bell state between the two qubits. These results pave the way for a novel
platform for quantum simulation
\cite{anandDualspeciesRydbergArray2024,MachuHybridCircularPlatform2026,nguyenQuantumSimulationCircular2018}
- and potentially metrology \cite{EcknerRydbergTweezerClock2023} - in which
both a data and an ancilla qubit are encoded within a single atom.

\section{Experiment}
\label{sec:experiment}

We begin our experiments by loading a tweezer array with $^{88}$Sr atoms from a two-stage
magneto-optical trap. After parity projection and ground state cooling, we
excite the atoms to the $\ket{79\text{F}, m=2}$ Rydberg state from where they
are promoted to the circular state $\ket{79\text{C}}$ via adiabatic rapid
passage. The details of our experimental setup have been described elsewhere
\cite{holzlMotionalGroundstateCooling2023a,holzlLongLivedCircularRydberg2024a,pultineviciusLonglivedGiantCircular2025a},
therefore in the following we will give only a short overview. At electric
fields of typically $E_z = \SIrange{1}{3}{\volt\per\cm}$ we can drive
microwave transitions to other states in the Rydberg manifold such as to the
$\ket{81\text{C}}$ via a two-photon transition (see
Fig.~\ref{fig:fig1}a).
Within our optically transparent black-body-radiation-suppression cavity
these states were shown to have lifetimes on the order of few milliseconds at
room temperature. For this work we define our circular Rydberg state (CRS)
qubit between $\ket{79\text{C}}=\ket{\downarrow}$ and
$\ket{81\text{C}}=\ket{\uparrow}$. State readout of the CRS qubit is achieved
via state-selective field ionization (SSFI) and subsequent time resolved ion
detection on a micro-channel plate (MCP) detector.

The long lifetimes in combination with the negligible autoionization rate of
the circular states allow for the optical manipulation of the ionic core. Its energy levels
are essentially those of $^{88}$Sr$^+$ (Fig.~\ref{fig:fig1}b),
giving rise to the dipole-allowed 5S$_{1/2}\rightarrow 5$P$_{1/2}$ and the narrow 5S$_{1/2}\rightarrow 4$D$_{5/2}$ quadrupole transitions.
The latter is extensively used as an optical frequency standard~\cite{lindvall88SrOpticalClock2025,marceauAbsoluteFrequencyMeasurement2025,akermanOperatingMultiIonClock2025},
optical qubit~\cite{wangDemonstrationQuantumLogic2010,katzQuantumLogicDetection2022,akermanUniversalGatesetTrappedion2015}, or part of coherent Rydberg excitation~\cite{higginsCoherentControlSingle2017a} in trapped-ion quantum experiments.
To address the ionic core clock transition inside the Rydberg atoms, we set up a Hz-linewidth diode laser system at \SI{674}{nm} (see Appendix~\ref{apdx:exp_setup} for details), entering
orthogonally to an applied magnetic field $\mathbf{k}\perp \mathbf{B}$ ($B_z\approx \SI{1.5}{G}$). Depending on the polarization, this allows us to address
all transitions with $\Delta m=\pm1, \pm2$~\cite{roos2000controllingIons} (see Fig.\ref{fig:fig1}f). To determine whether we have successfully shelved the ionic core
into the D-state, we apply a blow-out sequence consisting of a $\approx\SI{30}{\micro\second}$
`pushing' pulse on the \SI{422}{nm} 5S$_{1/2}\rightarrow 5$P$_{1/2}$, while repumping from the 4D$_{3/2}$ state at \SI{1092}{nm}.
Rydberg atoms that are shelved to the 4D$_{5/2}$ state do not scatter any photons and are subsequently detected via SSFI, while
those remaining in the S$_{1/2}$ state are pushed out of the detection region.

When adjusting the linear polarization of the clock laser between parallel and orthogonal to $\mathbf{B}$ and applying this shelving spectroscopy scheme we
identify eight resonances (Fig.~\ref{fig:fig1}e), corresponding to the allowed quadrupole transitions. On any of these transitions we
can drive coherent Rabi oscillations. In the following we define our optical
qubit between the $\ket{\text{S}}=\ket{\text{S}_{1/2}, m_J=-1/2}$ and
$\ket{\text{D}}=\ket{\text{D}_{5/2}, m_J=-5/2}$ states, for which we achieve a
typical Rabi rate of $\Omega\approx2\pi\times\SI{140}{kHz}$ (see inset of
Fig.~\ref{fig:fig1}e).

A more refined variation
of our detection scheme allows us to measure in the two-qubit
basis spanned by the CRS and optical qubit (see Fig.~\ref{fig:fig1}c). Here, we tune the push duration such that the atoms in
$\ket{\text{S}}$ are displaced towards the MCP detector. During the SSFI atoms in
$\ket{\uparrow}$ are ionized earlier than those in $\ket{\downarrow}$. Additionally atoms in $\ket{\text{D}}$
are accelerated over a longer distance than those in $\ket{\text{S}}$ and thus reach the MCP earlier.
In effect, this yields four distinct arrival time bins (Fig.~\ref{fig:fig1}d), one for each of the states
$\ket{\downarrow\text{S}},\ket{\uparrow\text{S}},\ket{\downarrow\text{D}}$, and $\ket{\uparrow\text{D}}$.

\section{Quadrupole interaction between the two qubits}
Having shown that we can individually manipulate the ionic core electron and use it to implement an optical qubit, we now consider
its interaction with the CRS qubit. The absence of exchange interaction and the resulting lack of autoionization
is the prerequisite for two independently addressable electronic qubits inside a single atom. However, the electrostatic interaction
between the quadrupole field of the CRS and the quadrupole moment of the 4D$_{5/2}$ levels leads to a coupling between the two qubits. In the following section we will
demonstrate that, using the established coherent control, we can characterize and tune the strength of this interaction.

\subsection{Measurement of the quadrupole moment}
\begin{figure}[t]
	\centering
	\includegraphics[width=\linewidth]{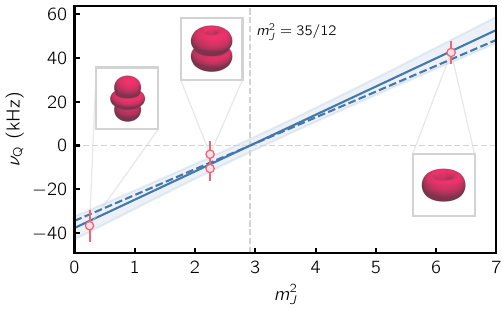}
	\caption{\textbf{Spectroscopic determination of the quadrupole moment $\Theta(\text{D}, {5/2})$.}
		The quadrupole shift $\nu_\text{Q}$ is shown as a function of the squared magnetic quantum number $m_J^2$ of the addressed D$_{5/2}$ level (red data).
		For the data presented, the Rydberg electron was prepared in the $\ket{79\text{C}}$ state and
		the quadrupole shift was determined by shelving spectroscopy of transitions with opposite
		relative magnetic moment, as described in the main text.
		From a linear fit with Eq.~\ref{eq:nu_QP} assuming the offset angle of $\vartheta=\SI{4.6}{\degree}$ calibrated in Sec.~\ref{sec:angle_diff} (blue solid line),
		we extract a quadrupole moment of $\Theta_\text{exp}=3.2(3)\,e a_0^2$, in agreement with the
		currently most precise theoretical and experimental values (blue dashed line) \cite{jiangElectricQuadrupoleMoments2008, shanivAtomicQuadrupoleMoment2016}.
		The insets show the core electron wave functions for the different values of $m_J$, ranging from prolate to oblate shapes.
		To remove an arbitrary offset the measured frequencies are shifted to cross zero for
		$m_J^2=35/12$ (vertical dashed line) by subtracting the fitted offset. Errorbars show $1\sigma$ confidence intervals.}
	\label{fig:fig2}
\end{figure}

The 4D$_{5/2}$ state differs from states with lower angular momentum in its permanent quadrupole moment $\Theta(\text{D},5/2)$ \cite{itanoExternalfieldShifts199Hg2000}.
In the presence of an electric quadrupole field this leads to an energy shift of its magnetic sublevels by
\begin{align}
	\label{eq:nu_QP}
	h\nu_\text{Q} = \frac{3}{40} \frac{\partial E}{\partial z} \,\Theta\left(\text{D}, 5/2\right) \left(\frac{35}{12} - m_J^2\right)(3\cos^2\vartheta -1),
\end{align}
with the magnetic quantum number $m_J$, the gradient $\partial E/\partial z$ of the quadrupole field along its principal axis,
and the angle $\vartheta$ between the magnetic field and the principal axis. In ion-trap based optical clock experiments this
shift is caused by the DC fields necessary to confine the ion and has been measured with excellent precision~\cite{barwoodMeasurementElectricQuadrupole2004,lindvallHighaccuracyDeterminationPaultrap2022,shanivAtomicQuadrupoleMoment2016}.
In the case of an ionic core inside a circular Rydberg atom the quadrupole field is generated by the Rydberg electron,
akin to the field created by a charged ring~\cite{muniOpticalCoherentManipulation2022,wirthQuadrupoleCouplingCircular2024}.
For a CRS the electric field gradient depends on the principal quantum number $n$
\begin{align}
	\label{eq:E_field_gradient}
	\frac{\partial E}{\partial z} = -\frac{4\, E_h}{e a_0^2} \frac{1}{4\,n^6-n^4},
\end{align}
with the Hartree Energy $E_h$, electron charge $e$ and Bohr radius $a_0$.
The resulting fields, $\partial E/\partial z\approx -\SI{40}{kV\per mm^2}$ for $n=79$, are large compared to those typically encountered in ion traps
leading to shifts on the order of tens of \si{kHz}. In this case $\nu_\text{Q}$ is negative (positive) for the more prolate (oblate) D-orbitals (see insets in Fig.~\ref{fig:fig2}).
We can exploit this to determine $\Theta(\text{D},5/2)$ by direct spectroscopy of the clock transition. To do so, we measure
the transition frequencies of transition pairs with equal $m_J^2$ and opposite relative magnetic moment
$\mu_\text{rel} =  \mu_\text{B} \left(m_{\text{D}5/2}\, g_{\text{D}5/2} - m_{\text{S}1/2} \,g_{\text{S}1/2}\right)$ using shelving spectroscopy as
described above with the tweezers switched off. As $\nu_\text{Q}$ is independent of the sign of $m_J$, taking the mean of transition frequencies with opposite $\mu_\text{rel}$ cancels out the Zeeman shift while preserving $\nu_\text{Q}$~\cite{dubeElectricQuadrupoleShift2005,dubeEvaluationSystematicShifts2013a}.
We show the resulting $\nu_\text{Q}$ as a function of $m_J^2$ in Fig.~\ref{fig:fig2}. A fit with Eq.~\ref{eq:nu_QP} yields $\Theta(\text{D},5/2)=3.2(3)\,e a_0^2$,
in good agreement with the currently most precise experimental, $\Theta_\text{exp}(\text{D},5/2)=2.973^{+0.026}_{-0.033}\,ea_0^2$
and theoretical, $\Theta_\text{theo}(\text{D},5/2)=2.935(17)\,ea_0^2$ , values~\cite{jiangElectricQuadrupoleMoments2008, shanivAtomicQuadrupoleMoment2016}.
Note that Eqs.~\ref{eq:nu_QP} and \ref{eq:E_field_gradient} also imply a differential quadrupole shift between circular states of different $n$, which is
\begin{align}
	\Delta\nu_\text{Q} = \nu_{\text{Q}, \downarrow} - \nu_{\text{Q}, \uparrow} \approx \SI{5.6}{\kilo\hertz}
\end{align}
for our choice of Rydberg and optical qubit states.

\subsection{Coherence of the optical qubit}
\begin{figure}[t]
	\centering
	\includegraphics[width=\linewidth]{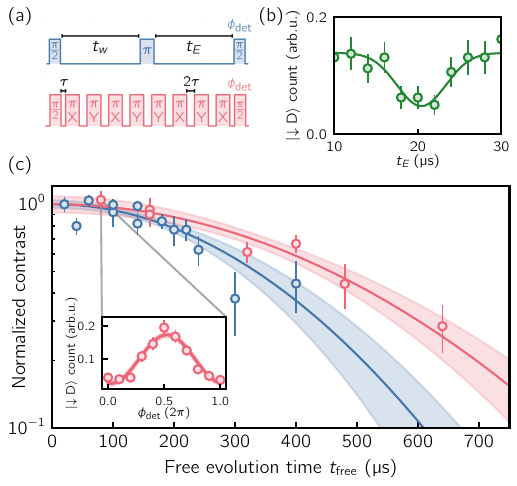}
	\caption{\textbf{Dynamical decoupling of the ion core optical qubit.} a) Schematic of the applied decoupling sequences spin-echo (blue) and XY8 (red). The pulse durations as well
		as the wait time $t_w$, echo time $t_E$ and XY8 time constant $\tau$ are not to scale. b) Representative measurement of the spin-echo envelope for zero detuning and $t_w=\SI{20}{\micro\second}$, characterizing the $\ket{\text{D}}\rightarrow\ket{\text{S}}$ transfer.
		From a Gaussian fit we deduce a reversible $T_2^*$ time of \SI{4.1(14)}{\micro\second}.
		c) Normalized
		fringe contrast for spin-echo (blue data)
		and XY8 (red data) decoupling as a function of the total free evolution time $t_\text{free}$, with
		$t_\text{free}=t_E+t_w$ for the spin-echo and $t_\text{free}=16\,\tau$ for the XY8 sequence. A fit with a Gaussian
		decay (solid lines) yields irreversible $T_2$ times of
		\SI{400(50)}{\micro\second} and \SI{550(50)}{\micro\second} for spin-echo and XY8, respectively. In the inset we show an exemplary Ramsey fringe obtained from scanning the closing pulse phase $\phi_\text{det}$ of the XY8 decoupling sequence. The
		errorbars and shaded regions represent $1\sigma$ confidence intervals.}
	\label{fig:fig3}
\end{figure}

We now want to turn towards the use of the states $\ket{\text{S}}$ and $\ket{\text{D}}$ as an optical qubit
inside the Rydberg atom. For the coherence of the two qubit system, the lifetime of the $\ket{\downarrow}$
state of $\tau\approx\SI{2}{ms}$ poses a rough upper limit, being much smaller than the lifetime
of the $\ket{\text{D}}$ state of \SI{0.4}{s}~\cite{madejSingleTrappedSr1990}. However, in our case other factors are more impactful. Most importantly, in our
current setup the optical tweezers are at a wavelength of \SI{540}{nm}, which is not magic for the
optical qubit but, in contrary, leads to anti-trapping of the $\ket{\text{D}}$ state. We thus need to
perform all interrogations of the optical qubit with the tweezers switched off. To measure the reversible dephasing
time $T_2^*$ of the optical qubit, we employ a resonant spin-echo sequence (with typical $\pi$-times of $T_\pi\approx\SIrange{3}{4}{\micro\second}$) as
sketched in Fig.~\ref{fig:fig3}a: $\pi/2$-pulse, wait time $t_w$, $\pi$-pulse, echo time $t_E$ and a closing $\pi/2$-pulse with variable phase $\phi_\text{det}$. With a wait time in the first arm of $t_w=\SI{20}{\micro\second}$
and varying $t_E$, the resulting echo centered around $t_E\approx t_w$ is shown in Fig.~\ref{fig:fig3}b). A fit with a Gaussian
envelope $\exp(-(t_E-t_w)^2/T_2^{*2})$ reveals $T_2^*=\SI{4.1(14)}{\micro\second}$, which is consistent with simulations of a
Doppler-broadened ensemble at a temperature of $T\approx\SI{10}{\micro\kelvin}$. The same dephasing would require magnetic field
fluctuations on the order of $\sigma_B\approx\SI{15}{mG}$, which is significantly larger than the magnetic-field stability
of our setup.

Since the Doppler shift is constant for freely expanding atoms it can be perfectly canceled out
by decoupling schemes such as spin-echo ($t_w=t_E$) or XY8 (see Fig.~\ref{fig:fig3}a). To measure the irreversible dephasing time $T_2$ under these two protocols,
we scan the phase of the closing $\pi/2$-pulse for different durations of the free evolution time. The resulting fringe contrast, normalized
by the average ion count is shown in Fig.~\ref{fig:fig3}c). To compare the decoupling schemes, we define the free evolution
time as $t_\text{free}=t_w+t_E$ for spin-echo and $t_\text{free}=16\,\tau$ for XY8, where $\tau$ is the XY8 time constant (see Fig.~\ref{fig:fig3}a).
In both cases the improvement over the $T_2^*$ time
is significant: from a fit with a Gaussian noise model $\propto\exp(-t_\text{free}^2/T_2^2)$ we extract $T_2=\SI{400(50)}{\micro\second}$
for spin-echo and $T_2=\SI{550(50)}{\micro\second}$ for the XY8 sequence.

\subsection{Differential quadrupole shift and angular dependence}
\label{sec:angle_diff}

\begin{figure}[t]
	\begin{center}
		\includegraphics[width=\linewidth]{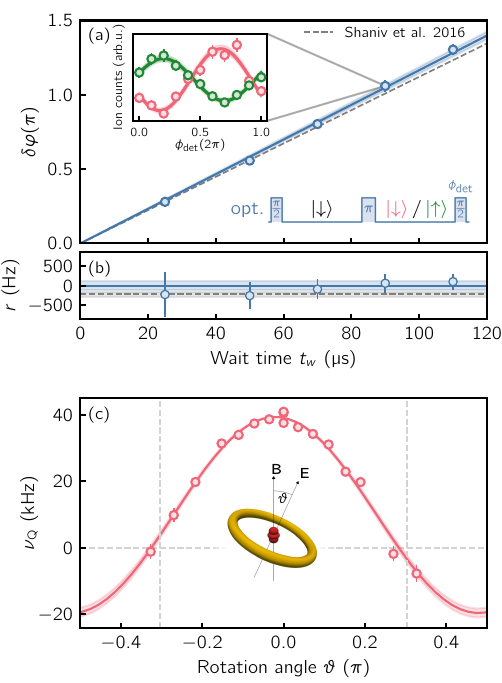}
	\end{center}
	\caption{\textbf{Spin-echo characterization of the quadrupole interaction.} (a) Measurement of the
		differential quadrupole shift between $\ket{\downarrow}$ and $\ket{\uparrow}$: We perform a
		spin-echo sequence on the optical clock qubit with the CRS qubit initially prepared in
		$\ket{\downarrow}$. The inset shows exemplary Ramsey fringes
		obtained for a wait time of $t_w=\SI{90}{\micro\second}$ when the CRS qubit is in $\ket{\uparrow}$ (green) or $\ket{\downarrow}$ (red) during the second interferometer arm. From the relative phase shift $\delta\varphi$ measured
		for different wait times $t_w$ (blue data) we obtain a
		differential quadrupole shift of $\Delta\nu_\text{Q}=\SI{5.83(11)}{\kilo\hertz}$ (blue line).
		The value of \SI{5.61(6)}{kHz}, calculated from \cite{shanivAtomicQuadrupoleMoment2016} is indicated by the grey dashed line. (b) We show the residuals $r$ of the fit. (c) Angular dependence of the quadrupole
		interaction: we vary the angle $\vartheta$ between the magnetic field $\mathbf{B}$ and the quadrupole field
		by aligning the circular Rydberg state along a strong, tilted electric field $\mathbf{E}$ in the
		second arm of the spin-echo sequence. The frequency shift $\nu_\text{Q}$, obtained from the
		phase shift of echo fringes after $t_w=\SI{20}{\micro\second}$, is shown in dependence on
		$\vartheta$. A fit with Eq.~\ref{eq:nu_QP} reveals a small initial angle between $\mathbf{B}$ and
		$\mathbf{E}$ of \SI{4.6(4)}{\degree} and yields a quadrupole moment of $\Theta(\text{D},5/2)=2.91(7)\,e a_0^2$. The vertical dashed lines mark the magic angle at which the quadrupole shift vanishes.
		All errorbars indicate $1\sigma$ confidence intervals and, where invisible, are
		smaller than the marker size.}
	\label{fig:fig4}
\end{figure}

The long coherence times enable us to perform more precise measurements on the combined system of
CRS and optical qubit. As already established, the interaction between the two is mediated via the
quadrupole field of the CRS at the location of the ionic core and takes the form of Eq.~\ref{eq:nu_QP}.
Crucially, in order to entangle the two qubits, one needs to resolve the differential
quadrupole shift $\Delta\nu_\text{Q}$, which is around a factor of ten smaller than the absolute quadrupole shift $\nu_\text{Q}$.

To measure $\Delta\nu_\text{Q}$, we employ the spin-echo measurement with $t_E = t_w$ on the optical qubit, but additionally perform a $\pi$-pulse
on the CRS qubit, overlapped with the optical $\pi$-pulse at the center of the interferometer, flipping it from $\ket{\downarrow}$ to $\ket{\uparrow}$. To rule out
systematic energy shifts of the optical qubit alone during the free time-of-flight, we perform a reference measurement
without the CRS $\pi$-pulse for every $t_w$. The relative phase shift $\delta\varphi=\varphi_{\uparrow}-\varphi_{\downarrow}$ corresponds
to the phase accumulated due to the differential quadrupole shift.
Further, we post-select for events in which the CRS is detected in the correct arrival time bin after the whole sequence.
This suppresses the contribution of elliptical states or other states in the Rydberg manifold that would
otherwise affect the measured phases $\varphi_{\uparrow/\downarrow}$. An example for the Ramsey fringes measured for $t_w=\SI{90}{\micro\second}$ is shown in Fig.~\ref{fig:fig4}a).
Note that the difference in fringe contrast is not caused by loss of coherence, but by a lower detection rate of $\ket{\uparrow}$ in multi-tweezer experiments.
From these measurements, we can extract the differential quadrupole shift as
\begin{align}
	\label{eq:nu_diff_phi}
	\Delta\nu_\text{Q} = \frac{\delta\varphi}{2\pi t_w}.
\end{align}
From a combined fit of data taken at different $t_w$ with Eq.~\ref{eq:nu_diff_phi}
we extract
\begin{align}
	\Delta\nu_\text{Q} = \SI{5.83(11)}{kHz},
\end{align}
corresponding to a quadrupole moment of $\Theta(\text{D}, 5/2)=3.09(6)e a_0^2$. The residuals of the fit are shown in Fig.~\ref{fig:fig4}b) alongside
the value calculated using $\Theta(\text{D}, 5/2)$ from Ref.~\cite{shanivAtomicQuadrupoleMoment2016} and Eq.~\ref{eq:E_field_gradient}.

The dependence of Eq.~\ref{eq:nu_QP} on the angle between $\mathbf{B}$ and the quadrupole field also yields an interesting
handle to tune the interaction between the CRS and the optical qubit. At our $n$ and typical $E$-fields of $>\SI{0.5}{\volt\per\centi\meter}$
the energy spacing defined by the linear Stark effect is more than an order of magnitude larger than that of the Zeeman effect. The
quantization axis of the CRS is thus given by $\mathbf{E}$~\cite{mehaignerieInteractingCircularRydberg2025} and can be rotated independently of that of the ionic core D-orbital, which aligns along $\mathbf{B}$.
We calibrate the orientation of $\mathbf{E}$ by spectroscopy of
the quadratically shifted $\ket{\downarrow}\rightarrow\ket{\uparrow}$
transition (see Appendix~\ref{apdx:field_calibration} for details) and define
$\vartheta$ with respect to $z$ (orthogonal to our suppression capacitor
plates, which we use to apply $E_z$). Then we perform a phase-sensitive
spin-echo measurement on the optical qubit with
$t_w=t_E=\SI{20}{\micro\second}$ and the CRS qubit in $\ket{\downarrow}$, where
we rotate $\mathbf{E}$ to a target $\vartheta$ in the second arm. Note that as
the linear Stark effect is on the order of tens of MHz, we can rotate the
circular orbital adiabatically within microseconds. The results of
these measurements are shown in Fig.~\ref{fig:fig4}c. We find good agreement of
our data with Eq.~\ref{eq:nu_QP} and demonstrate tunability of $\nu_\text{Q}$
over the whole range from its maximum to zero at the ``magic angle'' of $\vartheta\approx\SI{54.7}{\degree}$. The
measurement also reveals a small offset of \SI{4.6(4)}{\degree} between our
magnetic field and the $z$-axis. From the angular dependence, we deduce a
quadrupole moment of $\Theta(\text{D}, 5/2)=\SI{2.91(7)}{e a_0^2}$. Taking the
weighted mean of all three measurements performed in this work, we report
\begin{align}
	\Theta_\text{exp}(\text{D}, 5/2) = 3.02(5)_\text{stat} \,e a_0^2.
\end{align}
We estimate that the systematic effect of the applied electric field is at least an order
of magnitude below our spectroscopic resolution (see Appendix~\ref{apdx:stark_effect})

\section{Entanglement between CRS and optical qubit}
\begin{figure}[t]
	\begin{center}
		\includegraphics[width=\linewidth]{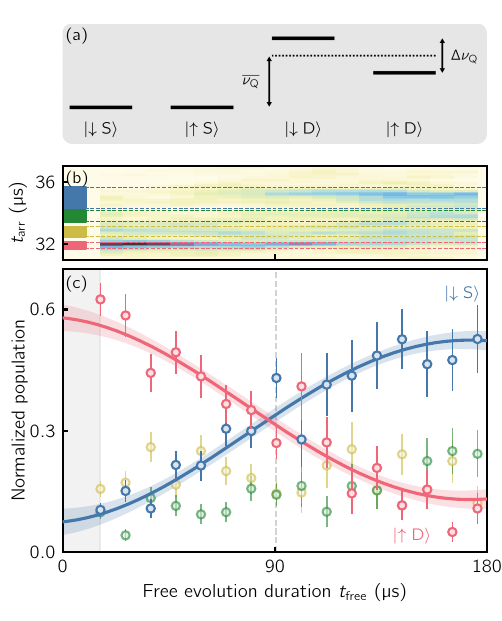}
	\end{center}
	\caption{\textbf{Dynamically decoupled two-qubit rotation between the optical and circular state qubit.} a) Schematic of the energy levels under the
		quadrupole interaction Hamiltonian Eq.~\ref{eq:H_QP}. The decoupling sequence on both qubits cancels out the mean quadrupole shift $\overline{\nu_\text{Q}}$, leading
		to a phase evolution with $\Delta\nu_\text{Q}$. The energy shifts are not to scale. b), c) We initially prepare the system in
		$\ket{\downarrow \text{S}}$ (blue data) and let it evolve while applying a parallel XY8 decoupling sequence on both
		individual qubits. Note that the net effect of this sequence without quadrupole interaction is a $\pi$ rotation on both qubits and
		thus we find the system in $\ket{\uparrow \text{D}}$ (red data) for short free evolution durations $t_\text{free}$. For longer durations, the
		two qubits rotate through the fully entangled state at $t_\text{free}\approx 1/2\Delta\nu_\text{Q}$ (grey dashed line) before the
		full population swap at $t_\text{free}\approx 1/\Delta\nu_\text{Q}$. Throughout we find around $\SIrange{10}{20}{\percent}$ population in $\ket{\downarrow \text{D}}$
		(green) and $\ket{\uparrow \text{S}}$ (yellow) due to our limited readout fidelity. We show an exemplary 2D arrival time histogram smoothed with a Gaussian kernel in (b) and the evaluated populations of several independent measurements in (c). The solid lines in (c) show a
		fit with a decaying sinusoidal function with fixed frequency $\Delta\nu_\text{Q}/2$ and free phase and coherence time. The greyed out area indicates durations that are technically inaccessible, due to pulse timing limitations. Errorbars indicate 1$\sigma$ confidence intervals.}
	\label{fig:fig5}
\end{figure}

\label{sec:entanglement_sequence}
Having established coherent control over both qubits and characterized the quadrupole
interaction between them, we now want to turn towards implementing a fundamental two-qubit gate.
The conceptually simplest way to do so would be a pulse on the clock transition, long enough
to spectroscopically resolve $\Delta\nu_\text{Q}$, thus implementing a CNOT gate. This would be
the fundamental ingredient for a non-destructive and local optical readout scheme of the CRS qubit.
To increase the robustness of the gate in a noisy environment however, it is favourable to implement it in a way that decouples
from (slow) variations of the environment.

In the co-rotating frame of the two unshifted qubits, the Hamiltonian describing the quadrupole
interaction takes the form
\begin{align}
	\label{eq:H_QP}
	\hat{H}_\text{Q} = \nu_{\text{Q}, \downarrow} \ket{\downarrow\text{D}}\bra{\downarrow\text{D}}  + \nu_{\text{Q}, \uparrow}  \ket{\uparrow\text{D}}\bra{\uparrow\text{D}},
\end{align}
where we can express the quadrupole shifts as $\nu_{\text{Q}, \downarrow/\uparrow} = \overline{\nu_\text{Q}} \pm \Delta\nu_\text{Q}/2$ (see Fig.~\ref{fig:fig5}a).
Our proposed two-qubit gate sequence consists of the time evolution according to Eq.~\ref{eq:H_QP} under a simultaneous decoupling scheme such as spin-echo or XY8 (see Appendix~\ref{sec:apdx_XY8} for details).
This increases the robustness against dephasing and removes the common mode energy shift $\overline{\nu_\text{Q}}$, giving rise
to a phase buildup proportional to $\Delta\nu_\text{Q}$ alone. The input state $\ket{\downarrow\text{S}}$ for example
evolves under a spin-echo sequence as
\begin{align}
	\ket{\downarrow\text{S}} \rightarrow \cos\left(\frac{\phi}{2}\right) \ket{\downarrow\text{S}} - i \sin\left(\frac{\phi}{2}\right)\ket{\uparrow\text{D}},
\end{align}
with $\phi = \pi\,\Delta\nu_\text{Q} \, t_\text{free}$. At $t_\text{free}=1/2\Delta\nu_\text{Q}\approx\SI{90}{\micro\second}$ the two qubits are in the fully entangled state $\ket{\downarrow\text{S}}-i\ket{\uparrow\text{D}}$.
Up to its direction of rotation, this operation is equivalent
to a two-qubit rotation $R_{yy}(-\phi)$ like it is realized by the original Mølmer-Sørensen gate~\cite{sorensenQuantumComputationIons1999}.
This means that together with single-qubit rotations our sequence forms a
universal gate set meaning that there exists e.g.\ a decomposition of CNOT into the decoupled
$R_{yy}$ sequence and fast single qubit rotations. Here we use spin-echo
decoupling as an instructive example, but the same principle applies for other
decoupling schemes such as XY8 which has the only additional effect of an $X$
gate on each qubit (see Appendix~\ref{sec:apdx_XY8}).

The experimental realization of this sequence is not without challenges. Unlike other systems,
where global decoupling pulses address two identical qubits~\cite{manovitzFastDynamicalDecoupling2017,baoDipolarSpinexchangeEntanglement2023,smithIndividuallyAddressedQuantum2024,nunnerichFastRobustLaserFree2025},
here we address a two-photon MW transition on one and an optical transition on the other qubit.
Our experimental routine is as follows: after preparation of the CRS qubit in $\ket{\downarrow}$, we perform a spin-polarization sequence, shelving the optical qubit
in $\ket{\text{D}}$, followed by a long push pulse and de-shelving back to $\ket{\text{S}}$. This removes the population in the undesired Zeeman sub-level of the S$_{1/2}$ state, that is
still present after Rydberg excitation from the $^{88}$Sr singlet ground state. We then perform a $\pi/2$-pulse, followed by an XY8 sequence and a closing $\pi/2$-pulse
with individually adjustable phase $\phi_\text{det}$ on both qubits. To stream the sequence of phase-coherent MW pulses with $\si{\micro\second}$ timing, we use a two-channel
vector signal generator. As we can drive the CRS qubit an order of magnitude faster than the optical qubit, we employ a pulse sequence where the MW $\pi$-pulses are much shorter than, and
mirror-symmetric with the respective optical $\pi$-pulses. Note that due to the two-photon nature of the MW transition, all programmed phases of MW pulses are halved compared to the optical pulses.
More details and numerical simulations of the XY8 sequence can be found in Appendix~\ref{sec:apdx_XY8}.
Finally, after the gate sequence we perform the two-qubit readout scheme described in Sec.~\ref{sec:experiment}.

In a first experiment, we vary the free evolution time $t_\text{free}=16\,\tau_\text{opt}$ (where $\tau_\text{opt}$ is the time constant of the optical XY8 sequence) while keeping the closing pulse phases
fixed at $\phi_\text{det}=0$. We initially prepare the system in the $\ket{\downarrow\text{S}}$ state.
The XY8 sequence for $\phi=0$ alone has the net effect of a $X$-gate on both qubits such that for short $t_\text{free}$, we
predominantly detect $\ket{\uparrow\text{D}}$ (see Fig.~\ref{fig:fig5}b and c). After a free evolution duration of $t_\text{free}\approx 1/\Delta\nu_\text{Q}\approx\SI{180}{\micro\second}$ the system
has evolved into the $\ket{\downarrow\text{S}}$ state, while the population of the other two states remains mostly constant.
A combined fit with decaying sinusoidal functions with frequency $f=\Delta\nu_\text{Q, theo}/2$ reveals good agreement with the expected $R_{yy}$ rotation,
apart from a small phase shift caused by finite pulse duration effects (see Appendix~\ref{sec:apdx_XY8}). From the fit we deduce
a fidelity of the full two-qubit rotation of $\mathcal{F}_{yy}=\SI{78(15)}{\percent}$, where SPAM errors are subtracted.

\begin{figure}[t]
	\begin{center}
		\includegraphics[width=\linewidth]{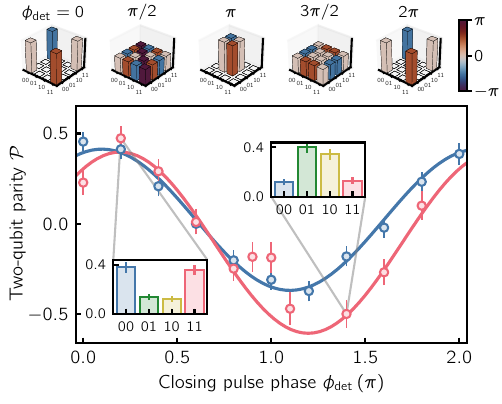}
	\end{center}
	\caption{\textbf{Parity of the final two-qubit state.}  At
		a free evolution duration of $t_\text{free}\approx1/2\Delta\nu_\text{Q}$, we scan the phase
		$\phi_\text{det}$ of the
		optical (blue) and microwave (red) closing pulse of the XY8 sequence. This reveals oscillations of
		the two-qubit parity $\mathcal{P}$ periodic in $\phi_\text{det}$. The insets show exemplary
		populations for the symmetric (left) and antisymmetric (right) state. At the top, we show density matrix representations
		of the final state
		resulting from simulations of the pulse sequence for different $\phi_\text{det}$. The absolute value of the complex matrix elements
		is encoded in the height of the bar, while the argument is color-coded.
		Here, we use the shorthand notation $00$ for $\ket{\downarrow\text{S}}$, $01$ for $\ket{\downarrow\text{D}}$, $10$ for $\ket{\uparrow\text{S}}$ and $11$ for $\ket{\uparrow\text{D}}$. Note that experimentally the phase of the
		microwave pulse is actually scanned by $\phi_\text{det}/2$ as we address a two-photon transition. Errorbars represent $1\sigma$
		confidence intervals.}
	\label{fig:fig6}
\end{figure}

Next we turn to investigate the entangled state prepared for $t_\text{free}\approx 1/2\Delta\nu_\text{Q}\approx\SI{90}{\micro\second}$.
As we show in Appendix~\ref{sec:apdx_XY8}, our gate sequence without the closing $\pi/2$-pulses results in the fully entangled
state $(\ket{\downarrow -} - \ket{\uparrow +})/\sqrt{2}$, where
$\ket{\pm}=(\ket{\text{S}}\pm\ket{\text{D}})/\sqrt{2}$. The final $\pi/2$-pulses rotate it such that (anti-)correlations become
apparent in the two-qubit basis accessible to our measurements. Varying the phase $\phi_\text{det}$ of either of the closing
pulses, while leaving the other at $\phi_\text{det}=0$, results in different final states which can show correlations ($\phi_\text{det}=0$), anti-correlations ($\phi_\text{det}=\pi$) or
no correlations ($\phi_\text{det}=\pi/2$) in our basis (see top of Fig.~\ref{fig:fig6}).
To characterize this behavior, we measure the populations $P_i$ and calculate the parity of the two qubit state
\begin{align}
	\mathcal{P} = P_{\downarrow\text{S}} + P_{\uparrow\text{D}} - P_{\downarrow \text{D}} - P_{\uparrow\text{S}},
\end{align}
while varying the phase of either the MW or optical closing pulse (see bottom of Fig.~\ref{fig:fig6}). For the exemplary phases $\phi_\text{det}$ explained above, the parity would be
$+1$ ($\phi_\text{det}=0$), $-1$ ($\phi_\text{det}=\pi$) or $0$ ($\phi_\text{det}=\pi/2$). Note that $\mathcal{P}$ alone is not a sufficient signature of a successful $R_{yy}$ operation:
in the hypothetical case of no quadrupole interaction
the two qubits would be in the separable $\ket{++}$ state before the closing pulse is applied. Scanning the phase of either of the closing pulses
would result in similar ``parity oscillations''. However, the resulting state would always be separable and not exhibit (anti-)correlations
in the two-qubit populations (the operation would just be $X\otimes X$ for $\phi_\text{det}=0$).
While this is not a quantitative  verification of entanglement creation, the data presented in Figs.~\ref{fig:fig5} and \ref{fig:fig6} exhibits
the expected time evolution under the $R_{yy}$ operation, as well as oscillations of $\mathcal{P}$ and shows the desired (anti-)correlations
in the two-qubit basis, consistent with the two-qubit entangled state. We find parity contrast of $0.8(1)$ and $1.0(1)$ when scanning the optical pulse and MW pulse phases respectively, largely limited
by overlap in our state detection. A phase difference of $0.10(3)\,\pi$ is  indicative of different dephasing mechanisms of the two qubits.

\section{Conclusion and Outlook}
We have demonstrated coherent control over an optical clock qubit in the ionic core
of a circular Rydberg atom. By first exciting one valence electron of neutral $^{88}$Sr to
a long-lived circular Rydberg state we could then independently address the S$_{1/2}\rightarrow$D$_{5/2}$ quadrupole
transition of the ionic core and detect successful excitation by rapidly scattering photons
on the ionic core D1-line. Adapting dynamical-decoupling schemes, we achieved $T_2$-times of the optical qubit
on the order of several hundred microseconds. Harvesting these coherence times, we characterized the electrostatic
quadrupole interaction between the circular Rydberg electron and the ionic-core D-orbitals. Here, we resolved the differential quadrupole
shift $\Delta\nu_\text{Q}$ between the $\ket{\downarrow}=\ket{79\text{C}}$ and the $\ket{\uparrow}=\ket{81\text{C}}$ circular Rydberg states
and showed that the quadrupole interaction can be rapidly tuned between zero and its maximum value by adjusting the orientation of the electric
quantization field of the CRS $\mathbf{E}$ with respect to the magnetic quantization field of the D-orbital $\mathbf{B}$.

Based on the quadrupole interaction, we proposed
and implemented a fundamental two-qubit gate between the CRS qubit and the ionic-core optical qubit: a two-qubit $R_{yy}$ rotation
akin to the operation implemented by the Mølmer-Sørensen gate. We found that we can perform a $R_{yy}(\pi)$ rotation with a fidelity of
$\mathcal{F}=\SI{78(15)}{\percent}$, and, characterizing the state resulting from a $R_{yy}(\pi/2)$ rotation found the
characteristic parity behavior and population correlations of an entangled state.

Our results open interesting pathways in the fields of quantum simulation and metrology. Coherent coupling between the
CRS and the optical qubit advertises the latter as an embedded ancilla qubit, while the former due to its long range interaction
with other CRS qubits can be used as a data qubit. From the $R_{yy}$ gate proposed here, together with single qubit
rotations, any other two-qubit gate can be implemented. A direct application of this would be non-destructive, local readout
of the CRS qubit by applying a CNOT gate, followed by fluorescence detection on the ionic-core D1-transition.
\textit{Vice versa} the clock laser can be used for site-resolved state preparation of the CRS array, either by
local addressing or by globally addressing all atoms while light-shifting selected ones out of resonance,
followed by mapping of the optical qubit state on the CRS qubit.

From the metrological perspective the embedded ionic-core offers the possibility of interrogating the clock transition in a microscopically
controlled environment and scaling the number of interrogated atoms up in a straight-forward manner.
Circular Rydberg state lifetimes of more than \SI{10}{ms} have been demonstrated~\cite{pultineviciusLonglivedGiantCircular2025a} and could potentially be further increased in cryogenic environments~\cite{jinExtendedRydbergLifetimes2026},
which could allow clock spectroscopy on timescales and at repetition rates used in state-of-the-art ion clock experiments~\cite{lindvall88SrOpticalClock2025}.

The current technical limitations in our experiment are measurement errors due to overlap of arrival time histograms
and anti-trapping of the D-state in our \SI{540}{nm} optical tweezers. A logical next step is fluorescence imaging
on the dipole allowed ionic-core S$_{1/2}\rightarrow$P$_{1/2}$ transition, which we expect to significantly improve readout fidelity
and allow for site-resolved detection. With improved readout fidelity at hand, a full characterization of the entangled state fidelity
of our $R_{yy}(-\pi/2)$ operation would be reasonable. The clock transition opens a path for a background free detection scheme
via excitation to the D$_{5/2}$ state, repumping to the P$_{3/2}$ state and fluorescence detection on the D2-line~\cite{blodgettNarrowlineElectricQuadrupole2025}.
Further, a tweezer trapping the optical clock qubit would allow to perform
experiments without switching off the confinement, thus enabling longer interrogation times. Magical trapping for the
$\ket{\text{S}}\rightarrow\ket{\text{D}}$ transition further facilitates the application
of sideband-resolved methods like cooling or thermometry.

\section*{Acknowledgments}
We are indebted to Tilman Pfau for invaluable support over the past years. We
thank Stephan Welte and the QRydDemo team for fruitful discussions. We
acknowledge funding from the Federal Ministry of Research, Technology and Space
under the Grants CiRQus and QRydDemo, the Horizon Europe Programme
HORIZON-CL4-2021-DIGITAL-EMERGING-01-30 via Project No. 101070144 (EuRyQa), and
the Carl-Zeiss foundation via IQST.

\section*{Contributions}
FT, AG and MT conducted the experiments and analyzed the data. FT, AG, MT, EP,
AH, and CH set up and maintained the experiment. FT performed the numerical
simulations. FT and FM wrote the manuscript with input from all authors. FM
supervised the project and acquired funding.

\section*{Data availability}
The data that support the findings of this study will be published on Zenodo upon publication of the manuscript.

\section*{Competing interest}
CH is founder and shareholder of Atomiq One GmbH.

\appendix

\section{Experimental setup and protocol}
\label{apdx:exp_setup}

In our setup we trap and cool strontium-88 atoms in a broad- (\SI{461}{nm}) followed by a narrow-linewidth (\SI{689}{nm})
magneto-optical trap (MOT), before transferring them to a linear chain of ten optical tweezers, spaced with $\Delta x = \SI{10}{\micro\meter}$.
The experiments presented in Figs.~\ref{fig:fig5} and \ref{fig:fig6}
were conducted using only a single tweezer to avoid the effect of interaction between CRS qubits.
Cooling and parity projection on the narrow \SI{689}{nm} $^1$S$_0\rightarrow ^3$P$_1$ transition leaves the atoms at $\approx\SI{4.2}{\micro\kelvin}$ with
an average filling of $\approx0.5$ atoms per tweezer~\cite{holzlMotionalGroundstateCooling2023a}. We subsequently use a resonance
enhanced three-photon excitation scheme to excite the atoms to the $\ket{n=79, l=3, m=2}$ state from which they are then transferred
to the circular state $\ket{n=79, l=78, m=78}$ via adiabatic rapid passage with an efficiency of $\approx\SI{70}{\percent}$~\cite{holzlLongLivedCircularRydberg2024a}.
To suppress the lifetime-shortening effect of black-body radiation (BBR) in the wavelength range relevant for single-photon transitions
between Rydberg states with $\Delta n=\pm1$, the atoms are held inside an optically transparent suppression cavity
extending their lifetime to $T_{1, \downarrow}=\SI{1.7(2)}{ms}$ and $T_{1,\uparrow}=\SI{3.8(3)}{ms}$ respectively~\cite{meinertIndiumTinOxide2020,pultineviciusLonglivedGiantCircular2025a}.
By applying a bias voltage to the electrodes of the cavity, we create an electric field of typically
$E_z = \SIrange{1}{3}{\volt\per\centi\meter}$ to orient the CRS. Additional ring electrodes allow us to apply
fields in the $x$-$y$-plane.
To address the microwave (MW) $\ket{\downarrow}\rightarrow\ket{\uparrow}$ transition in a fast and phase-coherent manner, we use
a vector signal generator (VSG) which can output pulses with programmable phase, amplitude and frequency and a minimum dead-time between pulses of \SI{6}{\micro\second}.
By using a conventional splitter/combiner and two phase-coherent output channels we reduce the dead-time between two arbitrary subsequent pulses to $\leq\SI{3}{\micro\second}$.

To address the optical clock transition of the ionic core, we use a clock laser system, consisting of a diode-laser seeded tapered amplifier
locked to a high-finesse ultra-low-expansion (ULE) cavity. We use a free-space single-pass acousto-optic modulator (AOM) for power stabilization, fast switching and phase-control and focus the beam
down using a cylindrical lens. The resulting waists of $w_z\approx\SI{60}{\micro\meter}$ and $w_y\approx\SI{350}{\micro\meter}$ at the location of the atoms allow for a more homogeneous
addressing of 2D tweezer arrays in the future with a peak intensity of \SI{0.8}{\watt\per\milli\meter^2}. We rotated the linear polarization of the clock laser
away from $\text{pol}\perp\mathbf{B}$ to detect all resonances shown in Fig.~\ref{fig:fig1} and
used $\text{pol}\perp\mathbf{B}$ in all other measurements to maximize the Rabi rate of the $\ket{\text{S}_{1/2}, m=-1/2}\rightarrow\ket{\text{D}_{5/2}, m=-5/2}$ transition.

To perform the XY8 two-qubit rotation sequence, we calibrate the Rabi rates of the optical qubit
as well as the CRS qubit for both VSG channels in independent measurements. Typical $\pi$-times for
our power settings are $T_{\pi, \text{MW}}\approx\SI{200}{ns}$ and $T_{\pi, \text{opt}}\approx\SIrange{3}{4}{\micro\second}$.
To correctly configure the $X$- and $Y$-pulses from different VSG channels, we measure the phase shift
induced by the MW combiner with a Ramsey measurement in which the opening and closing pulse are sourced from
different channels. Due to the much faster Rabi rates we can achieve on the CRS qubit, we adapt a sequence in which
all decoupling pulses are mirror-symmetric and $\tau_\text{opt}>\tau_\text{MW}$ (more details in Appendix~\ref{sec:apdx_XY8}).
For the typical $T_\pi$ given above the minimum $\tau_\text{opt}$ is the phase switching time of the AOM
driver of $\approx\SI{1}{\micro\second}$ (greyed out area in Fig.~\ref{fig:fig5}).

For the two-qubit readout sequence, we make use of the state-selective scattering on the ionic core D1
transition. We use a $\pi$-polarized pushing beam at \SI{422}{nm} propagating in the $x$-$y$-plane at a \SI{45}{\degree} angle with
the $x$-direction. The \SI{1092}{nm} repumping beam propagates along $x$ with $(\sigma_+ + \sigma_-)/2$ polarization in the atomic frame.
We take care to detune the \SI{422}{nm} from the two-photon resonance condition with the \SI{1092}{nm} laser and not saturate the
resulting dark resonances. The magnetic field of $B\approx\SI{1.5}{G}$ remixes the dark states in the D$_{3/2}$ manifold.
We apply a \SI{11}{\micro\second} pushing pulse and a waiting time of \SI{1}{ms} before SSFI
to sufficiently separate the arrival time histograms of the $\ket{\text{S}}$ and $\ket{\text{D}}$ bins.

\section{Angle dependence of the quadrupole interaction}
\label{apdx:field_calibration}

\begin{figure}
	\begin{center}
		\includegraphics[width=\linewidth]{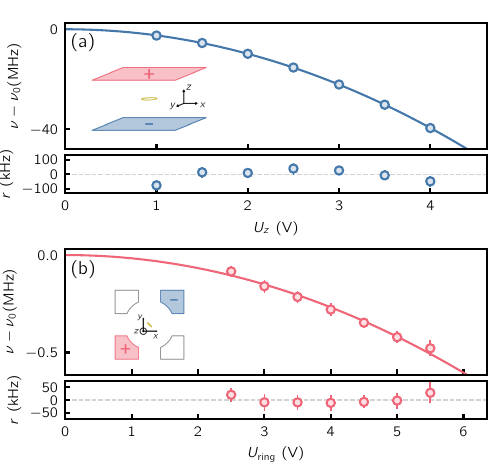}
	\end{center}
	\caption{\textbf{Calibration of electric fields.} We use MW spectroscopy on the quadratically shifted $\ket{79\text{C}}\rightarrow\ket{81\text{C}}$ transition
		to determine the scaling factor for rotation of $\mathbf{E}$. At a magnetic field of $B_z\approx \SI{1.6}{G}$ we apply a voltage to
		the $z$ electrodes (a) and to opposing ring-electrodes lying in the $x$-$y$-plane (b) and spectroscopically determine the
		shift of the transition from resonance. From fitting with a perturbative formula for the combined Stark- and Zeeman-shift (solid lines) we
		obtain effective scaling factors of $d_z = \SI{1.0115(6)}{cm}$ and $d_\text{ring}=\SI{12.2(6)}{cm}$. We also plot the residuals $r$ of the fits,
		errorbars indicate $1\sigma$ confidence intervals.}
	\label{fig:apdx_field_calibration}
\end{figure}
To calibrate the rotation angle shown in Fig.~\ref{fig:fig4}b),
we performed microwave spectroscopy of the quadratically Stark-shifted $\ket{79\text{C}}\rightarrow\ket{81\text{C}}$ transition (see Fig.~\ref{fig:apdx_field_calibration}).
For different voltages applied to the $z$ and ring-electrodes (in the $x$-$y$-plane), we measure the transition frequency
and fit the data with a perturbative expression taking into account Stark- and Zeeman-shifts~\cite{zimmermanStarkStructureRydberg1979}.
From this we obtain the scaling factors $d_i = U_i/E_i$ of
\begin{align*}
	d_z = \SI{1.0115(6)}{cm} \quad d_\text{ring} = \SI{12.2(6)}{cm},
\end{align*}
which we use to adjust the rotation angle with respect to the $z$-axis
\begin{align*}
	\vartheta = \arctan \left(\frac{U_\text{ring} d_z}{U_z d_\text{ring}}\right).
\end{align*}
We also  ensure that $|\mathbf{E}|=\text{const.}$ during the rotations.

For the data presented in Fig.~\ref{fig:fig4}b) we measured the Ramsey fringes for various $\vartheta$ and extracted
the fringe phase $\varphi$. Subtracting the maximum $\varphi$ yields the phase shift $\delta\varphi$ with respect to the maximum
$\nu_\text{Q}$ from which we obtain the frequency shift
\begin{align*}
	\delta\nu_\text{Q} (\vartheta) = \frac{\delta\varphi}{2\pi t_w},
\end{align*}
with the wait time $t_w=\SI{20}{\micro\second}$.
We fit the resulting $\delta\nu_\text{Q}$ with Eq.~\ref{eq:nu_QP}, leaving only $\Theta(\text{D}, 5/2)$ and $\vartheta_0$ (the angle between the $B$-field and $z$)
as open parameters. The data in Fig.~\ref{fig:fig4} are shifted up by the fitted magic angle detuning
$-\delta\nu_\text{Q}(\vartheta_\text{magic})$ for better readability.

\section{Systematic effect of external electric fields}
\label{apdx:stark_effect}

\begin{figure}
	\begin{center}
		\includegraphics[width=\linewidth]{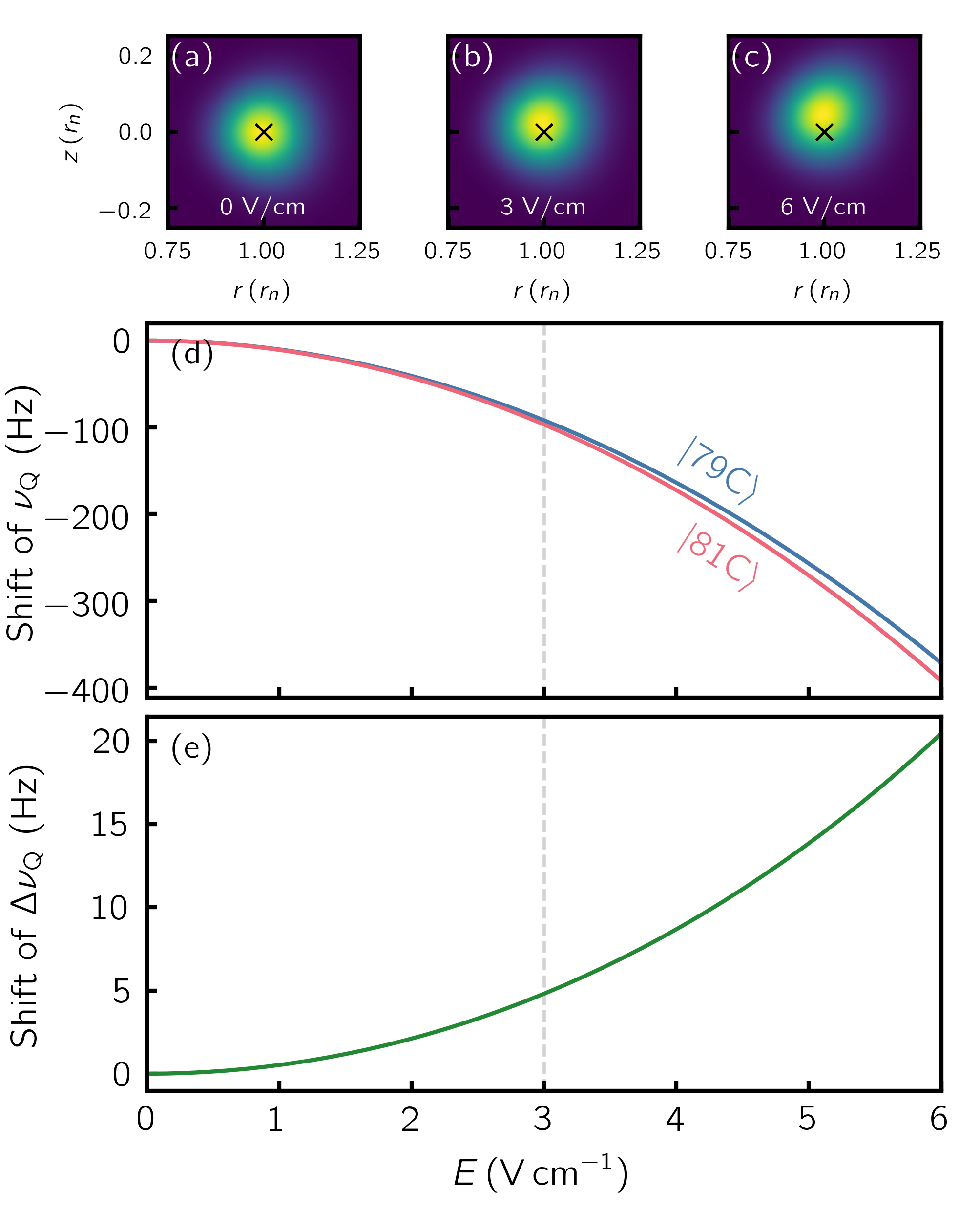}
	\end{center}
	\caption{\textbf{Systematic effect of the second order Stark shift.} The CRS is polarized by the strong electric quantization field along $z$. a)-c) We show the
		probability density of a CRS with $n=79$ in fields of $E = \SIrange{0}{6}{\volt\per\centi\meter}$. Effectively the $E$ field displaces the CRS along $z$, leading
		to a change of gradient at the location of the ionic core. This leads to a change of the quadrupole and differential quadrupole shifts (d,e), depending on the principal quantum
		number $n$ ($n=79$ in blue, $n=81$ in red). Note that while the absolute quadrupole shift decreases with $E$, the differential shift grows due to the difference in polarizability.
		The vertical dashed line indicates the maximum field applied throughout this work of $\approx \SI{3}{\volt\per\centi\meter}$,
		resulting in systematic shifts $<\SI{100}{\hertz}$ of $\nu_\text{Q}$ and $<\SI{5}{\hertz}$ of $\Delta\nu_\text{Q}$.}
	\label{fig:apdx_second_order_stark}
\end{figure}

Throughout our experiments we apply an electric field of $|\mathbf{E}|=\SIrange{1}{3}{\volt\per\centi\meter}$
as quantization field for the Rydberg states. As the circular states posses no permanent electric dipole moment,
they are Stark-shifted only in second order due to their polarizability. The corresponding displacement of the circular orbital out of the $x$-$y$-plane affects the quadrupole field at the location
of the ionic core. To quantify this effect we use the \texttt{pairinteraction} package~\cite{mogerleAccurateModelingRydberg2026}
to diagonalize the single Rydberg-atom Hamiltonian in an external electric field. We show the electron probability density of the
resulting state $\psi'$ around $r_n=a_0 n^2$ for $E_z=0,3,6\,\si{\volt\per\centi\meter}$ in Fig.~\ref{fig:apdx_second_order_stark}a-c), revealing visible polarization in larger
fields. We can use $\psi'$ to calculate the change in the quadrupole field experienced by the ionic core
\begin{align*}
	\frac{\partial E}{\partial z} = -2\braket{\psi'|\nabla \hat{E}^{(2)}_0|\psi'},
\end{align*}
where $\nabla E^{(2)}_0$ is the zeroth component of the electric field gradient tensor (its other components vanish due to cylindrical symmetry of the CRS)~\cite{itanoExternalfieldShifts199Hg2000}.
We can plug this into Eq.~\ref{eq:nu_QP} to obtain the variation of $\nu_\text{Q}$ and $\Delta\nu_\text{Q}$ (see Fig.~\ref{fig:apdx_second_order_stark}d-e).
The expected shifts of $\nu_\text{Q}$ and $\Delta\nu_\text{Q}$ at the maximum applied field of \SI{3}{\volt\per\centi\meter} are smaller
than \SI{100}{Hz} and \SI{5}{Hz}, respectively, and well below our current experimental resolution. We thus expect no systematic
effect of the applied $E$-field on our determination of $\Theta(\text{D}, 5/2)$.

\section{Details of the XY8 two-qubit rotation sequence}
\label{sec:apdx_XY8}

\begin{figure}[t]
	\begin{center}
		\includegraphics[width=\linewidth]{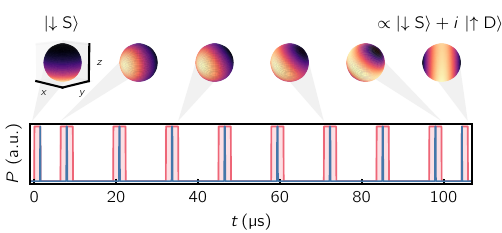}
	\end{center}
	\caption{\textbf{Husimi-$Q$-distribution during the XY8 sequence.} We show the normalized pulse sequence of MW (blue) and
		optical clock laser (red) for a two-qubit rotation to the fully entangled state ($t_\text{free}\approx1/2\Delta\nu_Q$). Here, we use experimentally realistic
		$T_\pi$ times of \SI{200}{\nano\second} for the CRS and \SI{3}{\micro\second} for the optical qubit. The upper plots show the Husimi-$Q$-distribution, plotted on the
		unit sphere, at different times. The buildup of entanglement is visible in the formation of a ring-like structure, the typical signature of triplet Bell states caused by their
		one-fold rotational invariance.}
	\label{fig:apdx_husimi}
\end{figure}

When the rotating frame Hamiltonian given in Eq.~\ref{eq:H_QP} is symmetrically decoupled with
a single ideal $\pi$-pulse, the effective time evolution becomes ($t=t_\text{free}$)
\begin{align}
	U_\text{eff}(t) & = e^{-i\hat{H}_\text{Q} t/2\hbar} (X\otimes X) e^{-i\hat{H}_\text{Q} t/2\hbar}                                                                                                                                           \\
	                & = -e^{-i\pi\nu_{\text{Q},\uparrow}t} \begin{pmatrix} 0 & 0 & 0 & 1 \\ 0 & 0 & e^{-i\pi\Delta\nu_\text{Q}t} & 0 \\ 0 & e^{-i\pi \Delta\nu_\text{Q}t} &0 & 0 \\ 1& 0 & 0 & 0  \end{pmatrix},
\end{align}
where $X$ is a $\pi$-rotation around $x$.
Wrapping this in a Ramsey sequence on both qubits and using $\phi_\text{Q} = \pi \Delta\nu_\text{Q} t$ we obtain
\begin{align*}
	U(\phi_\text{Q}) & = \left[R_x(\pi/2)\otimes R_x(\pi/2)\right] \\ &\times  \,U_\text{eff}(t)\,\left[R_x(\pi/2)\otimes R_x(\pi/2)\right] \\
	                 & =
	\begin{pmatrix}
		\cos\frac{\phi_\text{Q}}{2}   & 0                            & 0                            & -i\sin\frac{\phi_\text{Q}}{2} \\
		0                             & \cos\frac{\phi_\text{Q}}{2}  & i\sin\frac{\phi_\text{Q}}{2} & 0                             \\
		0                             & i\sin\frac{\phi_\text{Q}}{2} & \cos\frac{\phi_\text{Q}}{2}  & 0                             \\
		-i\sin\frac{\phi_\text{Q}}{2} & 0                            & 0                            & \cos\frac{\phi_\text{Q}}{2}
	\end{pmatrix},
\end{align*}
where we omitted a global phase. This operation is the two-qubit rotation $R_{yy}(-\phi_\text{Q})$ implemented by the
original Mølmer–Sørensen gate~\cite{sorensenQuantumComputationIons1999}.
Similarly, when using an XY8 decoupling between the Ramsey pulses and again neglecting global phases
the unitary evolution becomes
\begin{align*}
	U_{XY8}(\phi_\text{Q}) & =
	\begin{pmatrix}
		i\sin\frac{\phi_\text{Q}}{2} & 0                             & 0                             & -\cos\frac{\phi_\text{Q}}{2} \\
		0                            & -i\sin\frac{\phi_\text{Q}}{2} & -\cos\frac{\phi_\text{Q}}{2}  & 0                            \\
		0                            & -\cos\frac{\phi_\text{Q}}{2}  & -i\sin\frac{\phi_\text{Q}}{2} & 0                            \\
		-\cos\frac{\phi_\text{Q}}{2} & 0                             & 0                             & i\sin\frac{\phi_\text{Q}}{2}
	\end{pmatrix} \\
	                       & = (X\otimes X) R_{yy}(-\phi_\text{Q}),
\end{align*}
where the additional $X\otimes X$ reflects the even number of decoupling pulses. In this case $U_{XY8}$ conserves the parity $\mathcal{P}$
of the input state and e.g.\ for $\phi_\text{Q}=\pi/4$ transforms $\ket{\downarrow \text{S}}\rightarrow (\ket{\downarrow\text{S}}+i\ket{\uparrow\text{D}})/\sqrt{2}$. When switching the phase of either of the closing $
	\pi/2$-pulses to $\phi_\text{det}=\pi$, we instead obtain
\begin{align*}
	U_{XY8}'(\phi_\text{Q}) & =
	\begin{pmatrix}
		0                            & i\cos\frac{\phi_\text{Q}}{2} & -\sin\frac{\phi_\text{Q}}{2} & 0                            \\
		i\cos\frac{\phi_\text{Q}}{2} & 0                            & 0                            & \sin\frac{\phi_\text{Q}}{2}  \\
		\sin\frac{\phi_\text{Q}}{2}  & 0                            & 0                            & i\cos\frac{\phi_\text{Q}}{2} \\
		0                            & -\sin\frac{\phi_\text{Q}}{2} & i\cos\frac{\phi_\text{Q}}{2} & 0
	\end{pmatrix},
\end{align*}
inverting the parity $\mathcal{P}$ and transforming $\ket{\downarrow\text{S}}\rightarrow -(\ket{\downarrow\text{D}}-i\ket{\uparrow\text{S}})/\sqrt{2}$. We characterize this
behavior of our gate in Fig.~\ref{fig:fig6}.

Due to the faster Rabi rates possible on the CRS qubit, we have adapted an XY8 decoupling sequence with different pulse
durations between qubits, shown in Fig.~\ref{fig:apdx_husimi}.
We simulate the full pulse sequence, including driving Hamiltonians, numerically.
For an intuitive understanding of the buildup of entanglement during the gate sequence
implemented in Sec.~\ref{sec:entanglement_sequence}
we visualize the two-qubit state at different times using the Husimi-$Q$-distribution~\cite{husimiFormalPropertiesDensity1940}
\begin{align}
	\label{eq:husimi_Q}
	Q(\theta, \varphi) = \frac{3}{4\pi} \braket{\theta,\varphi|\rho}\braket{\rho|\theta,\varphi}
\end{align}
with the density matrix $\ket{\rho}\bra{\rho}$ and the coherent state
\begin{align}
	\ket{\theta,\varphi} = \bigotimes_{n}^{2} \cos\left(\frac{\theta}{2}\right)\ket{1}_n + e^{i\varphi}\sin\left(\frac{\theta}{2}\right)\ket{0}_n,
\end{align}
where the tensor product goes over the two qubit subspaces. We show the Husimi-$Q$ distribution at different steps of the XY8 sequence
at the example of the input state $\ket{\downarrow\text{S}}$ in Fig.~\ref{fig:apdx_husimi}.
In this case $t_\text{free}$ is chosen such that before the closing $\pi/2$-pulse, the system is in a fully entangled state $\propto\ket{\downarrow\text{S}}-\ket{\downarrow\text{D}}-\ket{\uparrow\text{S}}-\ket{\uparrow\text{D}}=\ket{\downarrow -}-\ket{\uparrow+}$. The closing pulse
rotates to our measured two-qubit basis and the sequence results in the fully entangled state $\left(\ket{\downarrow\text{S}}+i\ket{\uparrow\text{D}}\right)/\sqrt{2}$.
Over the course of the sequence the characteristic ring structure of triplet Bell states~\cite{leuchsTripletlikeCorrelationSymmetry2009} forms and is rotated to $z$
by the closing $\pi/2$-pulses.

\begin{figure}[t]
	\begin{center}
		\includegraphics[width=\linewidth]{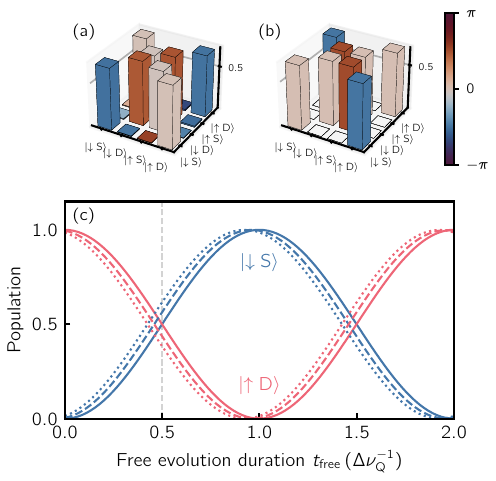}
	\end{center}
	\caption{\textbf{Simulation of the XY8 two-qubit rotation sequence.} We numerically simulate the XY8 two-qubit rotation sequence for $T_{\pi, \text{opt}}=\SI{3}{\micro\second}$,
		$T_{\pi, \text{MW}}=\SI{1}{\micro\second}$ and $t_\text{free}\approx1/2\Delta\nu_\text{Q}$ to
		reconstruct the effective gate matrix (a) and find that it matches that of an ideal $R_{yy}(-\pi/2)$
		operation depicted in (b) up to two single-qubit rotations $X$ and minor infidelities. Note that the two single-qubit $X$ gates are a consequence of the even number of decoupling pulses.
		The absolute values are encoded in the bar height, while the argument is color-coded. (c) We show the
		time-evolution of the two-qubit state populations for the initial state $\ket{\downarrow\text{S}}$ as a function of the free evolution time under XY8 decoupling. The population
		of the $\ket{\downarrow\text{S}}$ ($\ket{\uparrow\text{D}}$) state is shown in blue (red), the other states are not populated. For fast decoupling
		pulses, that satisfy $2\pi/\Omega\ll\tau$ (solid lines), the system evolves into the maximally entangled state for $t_\text{free}=1/2\Delta\nu_\text{Q}$ (vertical dashed line).
		Increasing the pulse durations on both qubits to $2\pi/\Omega\approx \tau$ (dashed lines), shifts the occurrence of maximal entanglement to earlier times, due to
		the mean interaction during the pulses. Slowing the decoupling of only one qubit (dotted lines) has a similar effect, as long as the pulses are symmetric.}
	\label{fig:apdx_gate_sim}
\end{figure}

The finite pulse durations lead to a deviation from the ideal gate operation, an effect that is known from quantum sensing and single qubit decoupling sequences~\cite{biercukExperimentalUhrigDynamical2009,ishikawaInfluenceDynamicalDecoupling2018}.
We quantify this effect by simulating the decoupling sequence for $\pi$-times of $[T_{\pi, \text{MW}}, T_{\pi, \text{opt}}] = [\SI{50}{ns}, \SI{50}{ns}], [\SI{3}{\micro\second}, \SI{3}{\micro\second}], [\SI{1}{\micro\second}, \SI{3}{\micro\second}]$ (see Fig.~\ref{fig:apdx_gate_sim}c),
finding an increasing phase shift of the $R_{yy}$ operation, due to the mean interaction during the decoupling pulses.
We find corresponding shifts of the entanglement time $t_\text{ent}$ of $t_\text{ent}-1/2\Delta\nu_\text{Q} \approx \SI{-2}{\micro\second}, \SI{-6}{\micro\second}, \SI{-14}{\micro\second}$,
in qualitative agreement with the experimentally observed value of $\SI{-3(4)}{\micro\second}$, although a measurement with
higher resolution is necessary to determine the precise experimental shift.

\clearpage

\bibliography{manually_added.bib}

\end{document}